# Nanolaser and logic gates on carbon nanotubes

I V Dzedolik[1] and S V Tomilin[1]

[1] V I Vernadsky Crimean Federal University, 4 Vernadsky Avenue, Simferopol, 295007, Russian Federation

E-mail: igor.dzedolik@cfuv.ru

**Abstract**

We consider theoretical models of the nanolaser and logic gates on carbon nanotubes (CNTs). In our work, it is shown at pumping the nanoresonator of the nanolaser on CNT by optical radiation using a quantum dot as nano light emitted diode (LED), the coherent flow of surface plasmon-polaritons arises when the generation threshold is exceeded. The coherent plasmonic flow from the partially reflecting mirror of the nanolaser enters the attached CNT plasmonic waveguide. Plasmonic logic gates NOT and OR based on CNTs represent the complete functional basis for binary logic. The proposed nanolaser and logic gates on CNTs can be used in plasmonic circuitry in the telecommunication frequency band.

## 1. Introduction

Modern trends towards miniaturization of computing and telecommunication systems create new requirements for the design and implementation of nanoscale devices [1-3]. Significant advances in recent years in nanostructure production technologies have made it possible to significantly reduce their size, including nanostructures operating at optical frequencies. Reducing the size of nanostructures not only helps reduce energy consumption, but also allows for an increase in the density of their placement on chips, which is especially important for high-tech systems where compactness and efficiency are required.

In connection with these trends, plasmonic circuitry is promising for application in various fields, such as nanoelectronics, nanophotonics, and nanoplasmonics. The element base of plasmonic circuitry is implemented on the basis of nanowaveguides, nanoresonators, plasmonic transistor [5,6] and logic gates [3,4,7-12]. Plasmonic circuitry currently, on the one hand, shows excellent results in the field of increasing the carrier frequencies of pulse signals up to optical ones [11,12], and clock frequencies in processors up to tens and hundreds of terahertz. But, on the other hand, metallic plasmonic structures as modulators, valves, and logic gates, in particular, based on the Mach-Zehnder interferometers [3,4,7-9,11,12], have relatively large sizes of tenths micrometers. Metallic plasmonic waveguides and elements of plasmonic circuits have large energy losses during heating and due to signal scattering on inhomogeneities. The relatively large size of the elements and losses due to heating hinder the development of plasmonic circuitry. In addition, the implementation of plasmonic circuitry also requires the presence of nanoscale radiation sources called nanolasers [1,2,13-32].

Currently, a large number of various nanolasers in the optical range have been developed and implemented, including spasers [13,15,25,27,28,31], generators based on graphene [19,23,31], on quantum wells [21], on semiconductor nanowires [17,20,21,24,29-31]. Terahertz range generators have been proposed also based on carbon nanotubes (CNTs) [14,22] and on a drift current [23]. The generation of the plasmonic flow instead of the photonic one in a nanolaser allows one to overcome the diffraction limit as a result of the exponential attenuation of the surface plasmon field, i.e. to implement a nanometer source of coherent radiation. Nanolasers have the ability to generate coherent electromagnetic radiation including optical radiation at subwavelength sizes.

CNTs, in which surface waves can be excited, have significant advantages over metal plasmonic elements: high conductivity and minimal heating losses, as well as much smaller transverse sizes of about 1-2 nm [33-39]. CNTs are single-layer or multilayer structures in the form of tubes, consisting of carbon atoms orderly located on the surface of the nanotube in the form of hexagons, as well as pentagons and heptagons in the bending region of the nanotube [36,37]. CNTs have both semiconductor and metallic properties depending on the type of crystal lattice called the chirality of CNT. When propagating ultrashort pulses of the optical range, CNTs exhibit nonlinear properties [40,41]. In addition, in CNTs the wavelength of surface modes is an order of magnitude shorter than in plasmonic waveguides when excited by an electromagnetic wave at a telecommunication frequency, which makes it possible to implement elements of plasmonic circuitry, including CNT-based nanolasers with minimal sizes.

Like any other laser, the nanolaser on CNT consists of three main elements: the nanoresonator, the gain medium, and the external energy source (pump). In produced nanolasers, the nanoresonators must provide a subwavelength size of the optical beam or plasmon flow and provide sufficient quality factor, which is necessary for generating coherent radiation at reasonable pumping levels [3]. A high-quality factor of the nanoresonator allows reducing the threshold values of pumping, which is important for the efficient operation of nanolasers in miniature devices. The higher the quality factor of the nanoresonator and the efficiency of the pumping, the lower the pumping threshold, which makes the systems more economical and efficient.

In our paper, the model of the nanolaser on CNT operating at a telecommunication frequency is proposed and theoretically investigated. In the nanolaser on CNT, we propose to use the CNT with zigzag chirality, which have semiconductor properties. Periodic surface inhomogeneities should be implemented at the ends of the nanotube, forming the Bragg gratings, i.e. such nanotube is the nanoresonator. It is possible to manufacture the nanoresonator with the Bragg gratings using CNT growing technologies by changing their growth regimes [38,39,42]. The reviews of the designs and properties of nanolasers as a new class of coherent generators are presented in the articles [27,28,31].

When pumping the nanoresonator of the nanolaser on CNT by optical radiation in continuous or pulsed mode, the coherent flow of plasmon-polaritons at the telecommunication frequency arises in it when the threshold is exceeded, and the nanoresonator length determines the frequency (wavelength). The flow of coherent plasmon-polaritons from the nanoresonator of the nanolaser can be directed into a plasmonic waveguide, including one based on CNT, connected to the nanoresonator. Coherent plasmon-polariton flow can be used in plasmonic logic gates. In addition, the nanolaser on CNT can be used as a signal amplifier in plasmonic circuitry.

The proposed laser on CNT is a generator of coherent radiation, including pulsed radiation, i.e. it can act as a clock generator in a plasmonic microcircuit. The advantage of the proposed nanolaser is generating a plasmonic pulse that is directed into the CNT plasmonic waveguide, and not into the surrounding space, i.e. no focusing system for the generated radiation is needed.

A semiconductor quantum dot (nano LED) is proposed as the pump source for the CNT nanolaser [32]. The emitting surface of the nano LED should cover the nanoresonator of CNT nanolaser within the Bragg gratings above the groove in the substrate. Voltage should be supplied to the nano LED using metal contacts (e.g. cobalt-molybdenum alloy, or others [39]). The semiconductor nano LED with the power (e.g. 3 pJ) and the radiation frequency (e.g. 650 nm) that is higher than telecommunication frequency, pumps the CNT nanolaser by flow as the spot size about 100-200 nm with linewidth near 10 nm. In this case, in the absence of given polarization of the LED radiation, both the TM mode and TE mode are excited in the nanoresonator of CNT nanolaser. If the radiation polarization of the pump LED is realized along the nanotube axis, then the TM mode is excited in the nanoresonator.

Also, in our work the schemes of plasmonic logic gates on CNTs are proposed. It is possible to create the fully plasmonic logic gate NOT as a result of propagating plasmonic signals on the nanotubes that is the different branches of the Mach-Zehnder interferometers. Plasmon logic gate OR can be realized on the basis of Y-splitter on CNTs. Such plasmonic structures can be created by means of known technologies from carbon nanotubes with diameter of 1-2 nm. The proposed logic gates NOT and OR on the basis of CNTs constitute a full functional basis for binary logic for plasmonic circuitry. The use of CNTs in elements of plasmonic circuitry will allow to significantly reduce the sizes of switching devices and processors operating on telecommunication frequencies.

The production of the proposed CNT nanolaser and logic gates can be implemented by method of growing nanotubes from the vapor phase of carbon in grooves of the given configuration and size in a dielectric substrate [39,42] on the upper surface of the substrate where the required grooves are etched. Carbon nanotubes are grown in the plane of the substrate. The Bragg gratings in the CNT nanotube can be realized by periodically changing the growth regimes of the nanotube to change its diameter.

A plasmonic waveguide of logic gate based on the CNT is connected to the output port of the nanolaser. In this case, the plasmonic pulse from nanolaser goes direct to the CNT-based logic gate, which should also receive a plasmonic signal from another gate. Using a combination of the proposed plasmonic CNT logic gates NOT and OR, the functional basis for binary logic of the plasmonic microcircuit is formed, i.e., the CNT logical circuitry is implemented, including plasmonic processors for various purposes [39].

The proposed CNT laser is working at the telecommunication frequency $\nu = 193.5 \cdot 10^{12} Hz$ (or higher). Thus, the clock frequency can be units or tens of terahertz of a plasmonic microcircuit with the proposed built-in nanolaser and logic gates, that is three orders of magnitude higher than the frequency of modern processors.

## 2. Model of nanolaser

In the proposed model of the nanolaser on CNT with semiconductor properties, the resonator (gain medium) for the nanolaser is the CNT with zigzag chirality at the ends of which there are periodic surface inhomogeneities (figure 1). The pumping of the nanolaser is carried out by radiation from a quantum dot in the optical range placed around the resonator (not shown).

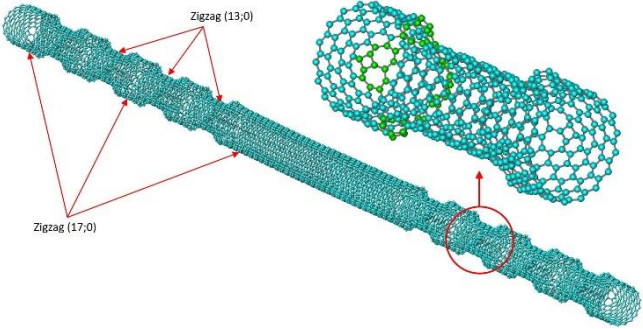

**Figure 1.** Sketch of the nanolaser on CNT.

The resonator of nanolaser is the CNT with zigzag chirality which has the periodic inhomogeneities of the boundary at the ends of nanotube represent the Bragg gratings as mirrors. CNT zigzag with the distance between neighbouring carbon atoms $d_0 = 0.142\ nm$ and chirality indices (17.0). The Bragg gratings are realized by periodically changing the diameter of the CNT without violating its chirality. Bragg gratings are at the end of the CNT with chirality indices (13.0) - (17.0).

Let us consider the propagation of modes along the CNT surface. The permittivities of the environment and the CNT are $\varepsilon$ and $\varepsilon_c$, the radius of the nanotube is $r_0$ [10]. The plasmon-polariton mode with the propagation constant $\beta$ falls on the boundary inhomogeneity (the Bragg grating). For maximum reflection from the Bragg grating, there should be constructive interference of waves $2n_{eff}d = \lambda/2$ upon reflection from a layer with the refractive index $n_{eff}$ and thickness $d$, i.e. the layer thickness should be $d = \lambda/4n$. In CNT, for modes in the boundary inhomogeneity we have $2\beta_h d = \pi$, i.e. the length of the CNT boundary inhomogeneity should be $d = \lambda_h/4$. When plasmon-polariton waves are reflected from the nanoresonator mirrors, the conditions for the emergence of a standing wave take place, i.e. multiple stimulated emission in CNT, which leads to the generation of the coherent plasmon-polariton flow, which is output into a plasmonic waveguide.

## 3. Plasmonic modes of carbon nanotube

Electromagnetic wave with the frequency $\omega$ interacts with the conduction electrons and bound electrons in a CNT. For a thin nanotube of radius $r_0$, the solutions of Maxwell's equations have the physical meaning for the components of electromagnetic field which tend to zero $r \to \infty$ with increasing distance from the nanotube axis $z$ if the thickness of nanotube is much smaller than the wavelength of the mode $r_0 \ll \lambda = 2\pi/\beta$. We can obtain the expression for the mode components of plasmon-polariton surface wave propagating along the CNT axis from the system of Maxwell's equations at the non-magnetic medium $\mu = 1$ with permittivity $\varepsilon$ for the monochromatic field $\sim \exp(-i\omega t)$ in the form of the Macdonald functions [10]

$$E_r = \frac{r_0^2}{w^2}\left[A\frac{-i\beta w}{r_0 K_l(w)}K'_l\left(\frac{wr}{r_0}\right) + B\frac{k_0}{K_l(w)}\frac{l}{r}K_l\left(\frac{wr}{r_0}\right)\right],$$
$$E_\varphi = \frac{r_0^2}{w^2}\left[A\frac{\beta}{K_l(w)}\frac{l}{r}K_l\left(\frac{wr}{r_0}\right) + B\frac{ik_0 w}{r_0 K_l(w)}K'_l\left(\frac{wr}{r_0}\right)\right],$$
$$H_r = \frac{r_0^2}{w^2}\left[B\frac{-i\beta w}{r_0 K_l(w)}K'_l\left(\frac{wr}{r_0}\right) - A\frac{k_0\varepsilon}{K_l(w)}\frac{l}{r}K_l\left(\frac{wr}{r_0}\right)\right], \quad (1)$$
$$H_\varphi = \frac{r_0^2}{w^2}\left[B\frac{\beta}{K_l(w)}\frac{l}{r}K_l\left(\frac{wr}{r_0}\right) - A\frac{ik_0\varepsilon w}{r_0 K_l(w)}K'_l\left(\frac{wr}{r_0}\right)\right],$$
$$E_z = A\frac{K_l(wr/r_0)}{K_l(w)}, \quad H_z = B\frac{K_l(wr/r_0)}{K_l(w)},$$

where the prime denotes the derivative of the function with respect to its argument, $w^2 = r_0^2(\beta^2 - k_0^2\varepsilon)$, $k_0 = \omega/c$. The dependence of the modes on time, azimuthal and longitudinal coordinates has the form $E_j, H_j \sim \exp(-i\omega t + il\varphi + i\beta z)$, $l = 0, \pm 1, \pm 2, \ldots$.

We find the components of modes with zero azimuthal index $l = 0$ from expressions (1) taking into account the property of the Macdonald function $K'_0 = -K_1$:
TM-mode
$$E_r = i\frac{\beta_0 r_0}{w}A\frac{K_1(wr/r_0)}{K_0(w)}, H_\varphi = i\frac{\varepsilon k_0 r_0}{w}A\frac{K_1(wr/r_0)}{K_0(w)}, E_z = A\frac{K_0(wr/r_0)}{K_0(w)}, \quad (2)$$

TE-mode
$$H_r = i\frac{\beta_0 r_0}{w}B\frac{K_1(wr/r_0)}{K_0(w)}, E_\varphi = -i\frac{k_0 r_0}{w}B\frac{K_1(wr/r_0)}{K_0(w)}, H_z = B\frac{K_0(wr/r_0)}{K_0(w)}, \quad (3)$$

where $\beta_0$ is the propagation constant of the corresponding mode.

## 4. Mode dispersion equation

Using the Leontovich boundary conditions $E_z = \zeta H_\varphi$ and $H_z = E_\varphi/\zeta$ on the nanotube surface at $r = r_0$, where $\zeta = \sqrt{\mu_c/\varepsilon_c}$ is the surface impedance of the conducting surface [43], we can obtain the dispersion equation for determining the propagation constants $\beta(\omega)$ of plasmon-polariton modes in CNT. Considering that $\mu_c = 1$ and $\zeta = 1/\sqrt{\varepsilon_c}$ for the CNT, we obtain the dispersion equation for the propagation constants $\beta_l(\omega)$ of plasmon-polariton modes of CNT from expressions (1) (Supplementary material)

$$K_l^2(w) + \varepsilon \frac{k_0^2 r_0^2}{w^2} K_l'^2(w) + i \frac{\varepsilon - \varepsilon_c}{\sqrt{\varepsilon_c}} \frac{k_0 r_0}{w} K_l(w) K_l'(w) = l^2 \frac{\beta^2 r_0^2}{w^4} K_l^2(w). \quad (4)$$

The argument of Macdonald function $w = r_0(\beta^2 - k_0^2 \varepsilon)^{1/2}$ must have a real value, i.e. $\beta > k_0 \varepsilon$, where the permittivity of the environment $\varepsilon$ has a real value.

For modes with zero azimuthal index $l = 0$, we obtain from equation (4) the dispersion equation in the form

$$K_0^2(w) + \varepsilon \frac{k_0^2 r_0^2}{w^2} K_1^2(w) - i \frac{\varepsilon - \varepsilon_c}{\sqrt{\varepsilon_c}} \frac{k_0 r_0}{w} K_0(w) K_1(w) = 0. \quad (5)$$

On the surface of the nanotube $r = r_0$, the modes with zero azimuthal index $l = 0$ are transformed into plane waves. It follows from the substitution into the expressions for modes (2) and (3) of the ratio $K_1(w)/K_0(w)$ obtained from the dispersion equation (5), where we need to take $\frac{K_1(w)}{K_0(w)} = -i \frac{w \sqrt{\varepsilon_c}}{\varepsilon k_0 r_0}$ for the TM mode, and $\frac{K_1(w)}{K_0(w)} = i \frac{w}{\sqrt{\varepsilon_c} k_0 r_0}$ for the TE mode. In this case, the amplitudes of the mode components have the form:

TM-mode
$$E_r = \frac{\beta_0 \sqrt{\varepsilon_c}}{\varepsilon k_0} A, \; H_\varphi = \sqrt{\varepsilon_c} A, \; E_z = A, \quad (6)$$

TE-mode
$$H_r = -\frac{\beta_0}{\sqrt{\varepsilon_c} k_0} B, \; E_\varphi = \frac{1}{\sqrt{\varepsilon_c}} B, \; H_z = B. \quad (7)$$

The mode propagation constants are determined from the ratio $K_1/K_0$, taking into account that $\sqrt{\varepsilon_c} = (\varepsilon_c'^2 + \varepsilon_c''^2)^{\frac{1}{4}} \exp(i\delta) = \sqrt{|\varepsilon_c|}(\cos\delta + i\sin\delta)$, where $\delta = \arctan(\varepsilon_c''/\varepsilon_c')/2$, $\varepsilon_c'$ and $\varepsilon_c''$ are the real and imaginary parts of the CNT permittivity. Then we obtain the dispersion equations for the TM mode and for the TE mode

$$wK_0(w) - \epsilon_{TM} K_1(w) = 0, \quad (8)$$
$$wK_0(w) - \epsilon_{TE} K_1(w) = 0, \quad (9)$$

where $\epsilon_{TM} = \frac{\varepsilon k_0 r_0}{\sqrt{|\varepsilon_c|}}$, $\epsilon_{TE} = \sqrt{|\varepsilon_c|} k_0 r_0$. We take into account that the imaginary part $i\epsilon_{TM,TE} \cos(\delta) = 0$ of the equations (8) and (9) is the real part and it is equal to zero, i.e. $\delta = \pi/2$, and $\sin(\delta) = 1$. We will estimate the volume $V$ of the mode field regarding the expression for the Macdonald functions $K_n \approx \sqrt{\pi r_0/2wr} \exp(-wr/r_0)$, assuming $r = (\beta^2 - k_0^2 \varepsilon)^{-1/2}$, then $V = \pi(r^2 - r_0^2)L$, where $L$ is the length of the resonator.

## 5. Resonator of nanolaser

The resonator of nanolaser (figure 1) is the CNT zigzag with periodic inhomogeneities of the boundary at the ends of the nanotube that are the Bragg gratings as mirrors. The Bragg gratings are realized by periodically changing the diameter of the CNT without violating its chirality. Let us consider the propagation of modes with the azimuthal index $l = 0$ along the surface of the CNT (expressions (6) and (7)).

The inhomogeneities of the CNT boundary are represented by $N$ periodic changes of the nanotube diameter, that are characterized by the matrix (Supplementary material)

$$M = \prod_{j=1}^{N} M_j = \begin{pmatrix} m_{11} & m_{12} \\ m_{21} & m_{22} \end{pmatrix}, \quad (10)$$

The amplitude reflection coefficients ρ have the form:
for TM-mode

$$\rho = \frac{E_{rr}}{E_{ri}} = \frac{m_{11} + m_{12}\frac{\varepsilon k_0}{\beta_0} - \frac{\beta_0}{\varepsilon k_0} m_{21} - m_{22}}{m_{11} + m_{12}\frac{\varepsilon k_0}{\beta_0} + \frac{\beta_0}{\varepsilon k_0} m_{21} + m_{22}}, \quad (11)$$

for TE-mode

$$\rho = \frac{E_{\varphi r}}{E_{\varphi i}} = \frac{m_{22} - \frac{k_0}{\beta_0} m_{21} - m_{11} + \frac{\beta_0}{k_0} m_{12}}{m_{11} - \frac{\beta_0}{k_0} m_{12} - \frac{k_0}{\beta_0} m_{21} + m_{22}}. \quad (12)$$

## 6. Coherent generation of nanolaser

For operation of the nanolaser in the coherent generation mode, the gain of radiation in one cycle (passage forward and backward between mirrors) must exceed the losses in the resonator and the laser radiation. The operating cycle of the nanolaser includes two successive reflections from the mirrors, which are the Bragg gratings (figure 1) with effective reflection coefficients $R_1$ and $R_2$ that include all losses. The attenuation of the flux is proportional to the product of the reflectivities $R_1 R_2$ on a path of length $2L$ in one cycle, where $L$ is the resonator length. The gain of the electromagnetic flux in the nanolaser in one cycle is equal to $I = I_0 R_1 R_2 e^{2\alpha L} = I_0 \exp[2\alpha L - \ln(R_1 R_2)]$ [44,46]. Generation occurs when $2\alpha L > |\ln(R_1 R_2)|$, i.e. the generation threshold of the nanolaser is $\alpha_0 = |\ln(R_1 R_2)|/2L$.

We define the quality factor of the nanolaser resonator as $Q = E/\delta E$, where $E = w_r S_r L$ is the energy stored in the resonator, $w_r$ is the energy density of the forward and backward flows, $S_r$ is the cross-sectional area of the resonator, $\delta E$ is the energy loss per cycle. The energy loss per cycle can be found as $\Delta E = \frac{1}{2} w_r S_r L[1 - \exp(-\ln(R_1 R_2))]$. The time of one cycle is equal to $t = 2L/v$, where $v$ is the wave velocity, and the period is $T_r = 2\pi/\omega$, then during one oscillation the energy loss is $\delta \tilde{E} = \frac{\Delta E}{2L/v} T_r = \frac{w_r S_r L v T_r}{4L} \left(1 - \frac{1}{\exp(\ln(R_1 R_2))}\right) \approx \frac{1}{4} w_r v S_r T_r |\ln(R_1 R_2)|$.

Then we find the quality factor of the CNT nanoresonator as $Q = \frac{N_{\lambda/2}}{|\ln(R_1 R_2)|}$, where $N_{\lambda/2} = 2L/(\lambda/2)$ is the number of half-waves of the standing wave in the nanoresonator, $\lambda = v T_r$ is the wavelength. Expressing the generation threshold through the quality factor of the nanoresonator, we obtain $\alpha_0 = N_{\lambda/2}/Q2L$ or $\alpha_0 \lambda/2 = Q^{-1}$. We determine the length of the nanoresonator by the number of half-waves of excited modes $L = (\lambda/2)n = (\pi/\beta)n$, where $n = 1,2,...$, and we find the minimum length of the nanoresonator taking into account the gain $\alpha \geq \alpha_0$ at $\alpha \lambda/2 \geq Q^{-1}$ as $L_{min} = 1/\alpha Q$.

## 7. Electronic spectrum of carbon nanotube

The band structure of CNT is determined by the dispersion equation [37]

$$\Delta E = \pm\gamma_{AB}\left\{1 + 4\cos\frac{k_y a}{2}\left[\cos\frac{\sqrt{3}k_x a}{2} + \cos\frac{k_y a}{2}\right]\right\}^{1/2}, \quad (13)$$

where $\gamma_{AB}$ is the overlap integral for the nearest atoms of the graphene lattice. The quantization condition of the electron wave vector in a CNT is represented as $\mathbf{R}\mathbf{k} = (m\mathbf{r}_1 + n\mathbf{r}_2)(\mathbf{1}_x k_x + \mathbf{1}_y k_y) = 2\pi s$, where $\mathbf{R} = m\mathbf{r}_1 + n\mathbf{r}_2$, $\mathbf{k} = \mathbf{1}_x k_x + \mathbf{1}_y k_y$ is the electron wave vector, $\mathbf{r}_1$ and $\mathbf{r}_2$ are the basis vectors of the graphene unit cell, $r_1 = r_1 = a$ is the graphene lattice constant, $(m,n)$ are the CNT chirality indices, $s = 1,2,3,...$ is an integer, numbering the allowed states of the electron. In a particular case, the CNT has a zigzag structure $(m,0)$, for which $k_y = \frac{2\pi}{\sqrt{3}d_0 m}$, $s = 1,2,3,...,m$, then we obtain the expressions for the half-width of the CNT band gap

$$\Delta E = \gamma_{AB}\left(1 + 4\cos\frac{\pi s}{m}\cos\frac{3k_x d_0}{2} + 4\cos^2\frac{\pi s}{m}\right)^{1/2}. \quad (14)$$

To generate a plasmon-polariton flow, the CNT must have semiconductor properties. In this case, the index $m$ of zigzag CNT must not be a multiple of three, i.e. the half-width of the forbidden zone $\Delta E \neq 0$ must not be zero [35,37]. The width of the forbidden zone is determined by expression (14) $E_g = 2\Delta E$ in the zigzag CNT with semiconductor properties with indices $(m,0)$ and the diameter $D = \frac{\sqrt{3}d_0}{\pi}m$, where $d_0 = 0.142\ nm$ is the distance between neighboring carbon atoms in the graphite plane.

We assume that the vector $k_x$ changes in the first Brillouin zone $k_x = \left(-\frac{\pi}{T}, \frac{\pi}{T}\right)$, where $T = a\sqrt{t_1^2 + t_2^2}$ is the modulus of the translation vector $\mathbf{T} = t_1\mathbf{r}_1 + t_2\mathbf{r}_2$ (it is perpendicular to the vector $\mathbf{R}$), $t_1 = \frac{2n+m}{d}$, $t_2 = -\frac{2m+n}{d}$, $d$ is the greatest common divisor for $2n + m$ and $2m + n$, $a = \sqrt{3}d_0$ [32]. The electronic spectrum of the zigzag CNT with indices (17,0) is shown in figure 2. From the analysis of the graphs in figure 2 it follows that the forbidden band in the zigzag CNT with indices (17,0) is determined by the difference in energies of the valence band and the conduction band $2\Delta E$ of the eleventh subband $s_{11}$. This energy transition can be used in the CNT laser resonator of appropriate length.

## 8. Interband electron transitions

If an electron in the CNT with semiconductor properties interacts with the electromagnetic pump field, then the probability of the electron transition from energy level $n$ to level $n'$ per unit time is equal to $w_{nn'} = \frac{2\pi}{\hbar}|\hat{H}'_{nn'}|^2\delta(E_n - E_{n'} - \hbar\omega)$ [46]. The Hamiltonian of the perturbation during the interaction of the electron with the electromagnetic field can be represented in the form $\hat{H}' = \frac{e}{2}\mathbf{r}\mathbf{E}_a\left(e^{i\omega t - i\mathbf{k}_{opt}\mathbf{r}}\right) + c.c.$, where $\mathbf{E}_a = const$ is the amplitude of the electric field. In this case, the matrix element of the perturbation operator is $\hat{H}'_{vc} = \frac{e}{2}\int d^3r\,\psi^*_{vk}(\mathbf{r})(\mathbf{r}\mathbf{E}_a)\psi_{ck'}(r)e^{i(\mathbf{k}'-\mathbf{k}-\mathbf{k}_{opt})\mathbf{r}}$. The value of this integral tends to zero due to the oscillating factor, except for the case $\mathbf{k}' - \mathbf{k} = \mathbf{k}_{opt}$.

In the parabolic approximation for the subbands of the valence band and the conduction band (figure 2) under the condition of a direct interband transition $k' = k \equiv k_x$, the electron energy is equal to the difference in the band energies plus the energy of the forbidden band $E_c - E_v = \frac{\hbar^2 k^2}{2}\left(\frac{1}{m_c} + \frac{1}{m_v}\right) + E_g$ [46]. In this case, the probability of an electron transition from the valence band to the conduction band per unit time is equal to $w_{vc} = \frac{2\pi}{\hbar}|\hat{H}'_{vc}|^2\delta\left(\frac{\hbar^2 k^2}{2m_{vc}} + E_g - \hbar\omega\right)$, where $m_{vc} = m_v m_c/(m_v + m_c)$ is the reduced effective mass of an electron.

The number of electron interband transitions $N = \int_0^\infty dk\,w_{vc}\rho(k)$ at the density of electron states $\rho(k) = 2 \cdot \frac{4\pi k^2}{8\pi^3}V = \frac{k^2 V}{\pi^2}$ for a body of volume V is equal to (Supplementary material)

$$N = \frac{V(2m_{vc})^{3/2}}{\hbar^4\pi}|\hat{H}'_{vc}|^2\sqrt{\hbar\omega - E_g}. \quad (15)$$

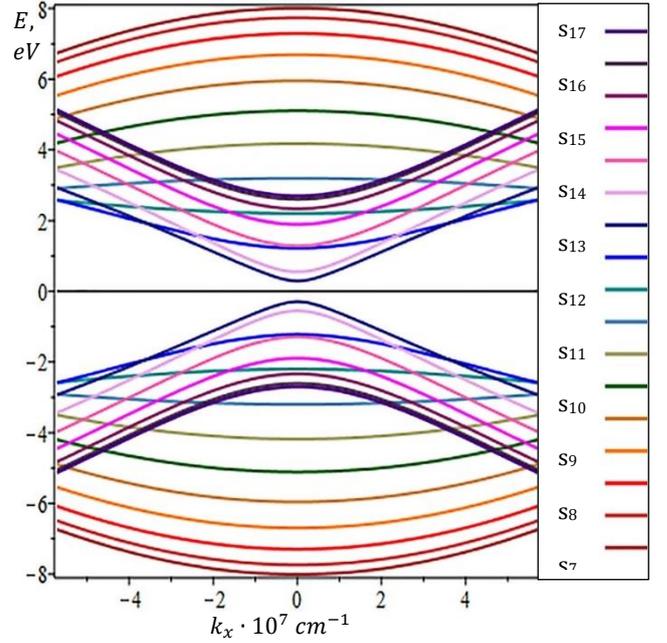

**Figure 2.** Electron spectrum of CNT zigzag with indices (17,0), number of subbands $s = 1,2,3,...,17$.

## 9. Carbon nanotube permittivity

The permittivity of the CNT is represented as [47] (Supplementary material)

$$\varepsilon_c(\omega) = 1 + \frac{2m\omega_e^2}{\hbar}\sum_n |r_{n0}|^2 \frac{\omega_{n0} - i\Gamma_n/2}{(\omega_{n0} - i\Gamma_n/2)^2 - \omega^2}, \quad (16)$$

where $\omega_e^2 = 4\pi e^2 N_e/m_{eff}$ is the square of electron plasma frequency, $m_{eff}$ is the effective mass of the electron, $N_e$ is the number of electrons per unit volume. We assume that the plasmon-polariton modes (expressions (1), (2) and (3)) are excited at the frequency of the direct electron transition $\omega = \omega_{nn'}$ in the CNT. Then the permittivity of the CNT during emission (absorption) of the photon at the frequency of electron transition $\omega$ is equal to

$$\varepsilon_c(\omega) = 1 + 12\pi c^3 N_e \left[\frac{4\Gamma}{\omega^2(16\omega^2+\Gamma^2)} + \frac{i}{\omega^3}\left(1 + \frac{\Gamma^2}{16\omega^2+\Gamma^2}\right)\right]. \quad (17)$$

## 10. Absorption coefficient

The absorption coefficient of a CNT is defined as the ratio of the absorbed power per unit volume $P_a = N\hbar\omega/V$ to the average power density of the incident electromagnetic radiation flux passing through a unit area $\alpha(\omega) = N\hbar\omega/|\overline{\mathbf{S}}|V$, where $|\overline{\mathbf{S}}| = \frac{c}{8\pi}\overline{|\mathbf{E} \times \mathbf{H}|}$ is the time-averaged Poynting vector (Supplementary material). For CNT modes with zero azimuthal index $l = 0$, we find the absorption coefficients for the TM mode and TE mode by substituting the number of interband transitions (15),

$$\alpha_{TM}(\omega) = \frac{(2m_{vc})^{3/2}}{\hbar^3\pi}\frac{|\hat{H}'_{vc}|^2}{(c/4\pi)S_{TM}A^2}\omega\sqrt{\hbar\omega-E_g}, \quad (18)$$

$$\alpha_{TE}(\omega) = \frac{(2m_{vc})^{3/2}}{\hbar^3\pi}\frac{|\hat{H}'_{vc}|^2}{(c/4\pi)S_{TE}B^2}\omega\sqrt{\hbar\omega-E_g}, \quad (19)$$

where $S_{TM} = \sqrt{\beta_0^2/k_0^2 + \varepsilon}$, $S_{TE} = |\varepsilon|^{-1}\sqrt{\beta_0^2/k_0^2 + \varepsilon}$, $\beta_0$ is the propagation constant of the TM mode or TE mode, respectively. Substituting the squares of the matrix elements into expressions (18) and (19) (Supplementary material), we obtain the absorption coefficients for the TM mode and TE mode

$$\alpha_{TM}(\omega) = \frac{e^2(2m_{vc})^{3/2}}{c\hbar^3\sqrt{\beta_0^2/k_0^2+\varepsilon}}|y_{vc}|^2\omega\sqrt{\hbar\omega-E_g}, \quad (20)$$

$$\alpha_{TE}(\omega) = \frac{e^2(2m_{vc})^{3/2}}{c\hbar^3\sqrt{\beta_0^2/k_0^2+\varepsilon}}r_0^2|\varphi_{vc}|^2\omega\sqrt{\hbar\omega-E_g}. \quad (21)$$

## 11. Gain factor of nanolaser

The conduction band of the CNT is partially filled with electrons up to the Fermi level $E_{Fc}$, and the valence band is empty up to the Fermi level $E_{Fv}$ at temperature $T\neq 0$. Electrons from the conduction band pass to the valence band with the emission of a photon in the energy range $E_g < \hbar\omega < E_{Fc} - E_{Fv}$, (where $E_{Fv} < 0$), and the absorption coefficients (20) and (21) $I = I_0 R_1 R_2 e^{2\alpha L}$ change sign from minus to plus [46]. In this case, the plasmon-polariton wave is amplified, i.e. the generation of CNT nanolaser takes place.

If the photon energy of the external electromagnetic field (pump) is $\hbar\omega_p < E_g$, then the absorption coefficient is $\alpha_{TM,TE}(\omega) = 0$, also the absorption occurs $\alpha_{TM,TE}(\omega) < 0$ at the pump photon energy $\hbar\omega_p > E_{Fc} - E_{Fv}$. Thus, during pumping the inversion of energy state occurs in the system if the energy of the pumping photons belongs to the range $\hbar\omega_p > E_{Fc} - E_{Fv}$, and photons with energy $E_g < \hbar\omega < E_{Fc} - E_{Fv}$ are emitted by the inverted CNT medium.

The expression for the gain factors (20) and (21) includes the squares of the matrix elements of the dipole perturbation Hamiltonian for the TM mode and TE mode, which can be determined through the relaxation frequency $\Gamma = A_{mn}$ (Supplementary material). Then we obtain the gain factor of the nanolaser in the form

$$\alpha(\omega) = \frac{3c^2(2m_{vc})^{3/2}\Gamma}{2\sqrt{\beta_0^2/k_0^2+\varepsilon}}\frac{\sqrt{\hbar\omega-E_g}}{\hbar^2\omega^2}, \quad (22)$$

where $\beta_0$ is the propagation constant of the TM mode or TE mode of the CNT, respectively.

## 12. Nanolaser parameters

We assume that the nanolaser generation occurs at the telecommunication frequency $\omega = 2\pi c/\lambda_0 = 1.216 \cdot 10^{15} s^{-1}$ ($\lambda_0 = 1.55\ \mu m$ in air, $0.8\ eV$). If we take the value of the overlap integral $\gamma_{AB} = 2.7\ eV$ in the dispersion equation (14), and the number of electrons per unit volume of the zigzag CNT $N_e = 1.8 \cdot 10^{12}\ cm^{-3}$ in expression (17), then the permittivity of the CNT for the frequency under consideration has the value $\varepsilon_c = 1.0 + i1.0$. For the given electron concentration in the conduction band of CNT, relaxation frequency $\Gamma = \frac{2e^6\omega}{3\hbar^3 c^3} = 3.13 \cdot 10^8\ s^{-1}$, such permittivity of CNT $\varepsilon_c$ and the permittivity of environment surrounding CNT $\varepsilon = 2.09$, for the zigzag CNT with indices (17,0) with the radius $r_0 = 0.665\ nm$ for modes with zero azimuthal indices (expressions (8) and (9)) we find the propagation constants of TM mode $\beta_{0TM} = 5.62 \cdot 10^5 cm^{-1}$ ($\lambda_{TM} = 112\ nm$) and of TE mode $\beta_{0TE} = 4.51 \cdot 10^5 cm^{-1}$ ($\lambda_{TE} = 140\ nm$).

The CNT zigzag with indices (17,0) has the radius $r_0 = 0.665\ nm$, the CNT boundary valleys representing the Bragg gratings have the radius $r_h = 0.509$ nm (CNT indices (13,0)). For the "high reflector" (left Bragg grating, figure 1) with the repeating of the CNT valleys $n_1 = 4$ times, and for the "output coupler" (right Bragg grating) with the repeating $n_2 = 4$ times, the reflection coefficients of the nanoresonator by intensity are $R_1 = \rho_{(n_1)}\rho^*_{(n_1)}$ and $R_2 = \rho_{(n_2)}\rho^*_{(n_2)}$.

The mode propagation constants for CNT (13.0) are $\beta_{0TM} = 6.27 \cdot 10^5 cm^{-1}$ and $\beta_{0TE} = 5.04 \cdot 10^5 cm^{-1}$, the reflection coefficients are $R_{1TM} = R_{2TM} = 0.173$ and $R_{1TE} = R_{2TE} = 0.175$, the quality factor is $Q_{TM} = 10$ and $Q_{TE} = 8$, the threshold generation coefficients are $\alpha_{0TM} = 1.76 \cdot 10^4 cm^{-1}$ and $\alpha_{0TE} = 1.74 \cdot 10^4 cm^{-1}$.

We find the value of the band gap in the zigzag CNT equal to $E_g = 0.586\ eV$ from expression (14) at $k_x = 0$. Also we determine the reduced effective mass of the electron $m_{vc} = m_v m_c/(m_v + m_c)$ for the values of the electron wave vector $k = 0.51 \cdot 10^7\ cm^{-1}$ for a direct interband transition at the frequency of $\omega = 1.216 \cdot 10^{15} s^{-1}$ ($0.8\ eV$). We find $m_v = m_c = (d^2\Delta E/dp_x^2)^{-1}$, where $p_x = \hbar k_x$, and obtain the value of the reduced mass $m_{vc} = m_c/2 = 0.91 \cdot 10^{-28}\ g$ at the generation frequency taking into account the symmetry of the spectral branches (figure 2).

Substituting the calculated parameters into expression (22), we find the values of the nanolaser gain coefficients for the TM mode $\alpha_{TM} = 5.10 \cdot 10^4 cm^{-1}$ and for the TE mode $\alpha_{TE} = 6.34 \cdot 10^4 cm^{-1}$, which exceed the threshold values of generation for the nanoresonator on the zigzag CNT for the TM mode by 2.9 times and for the TE mode by 3.6 times.

For the nanoresonator of minimum length with the Bragg gratings $L_{TM} = 592\ nm$ (mode field volume $V_{TM} = 5.94 \cdot 10^5 nm^3$) during generation in the TM mode, the Purcell coefficient is $F_{TM} = \frac{3\lambda_{TM}^3}{4\pi^2}\frac{Q}{V} = 1.79$. When the generation threshold for the plasmon-polariton flux intensity is exceeded, the CNT nanolaser generates coherent radiation in the TM mode with the frequency $\omega = 2\pi c/\lambda_0 = 1.216 \cdot 10^{15} s^{-1}$ and the linewidth $\Delta\omega_r = F_{TM}\Gamma = 5.59 \cdot 10^8\ s^{-1}$. When generating is in the TE mode, the nanolaser has the following parameters: $L_{TE} = 737\ nm$ ($V_{TE} = 11.6 \cdot 10^5 nm^3$), $F_{TE} = 1.42$, $\Delta\omega_r = 4.45 \cdot 10^8\ s^{-1}$. In the considered model of the CNT nanolaser, the generation linewidth is less than the angstrom without taking into account additional factors of linewidth broadening.

## 13. Logic gates

All CNTs in proposed plasmonic logic gates have metallic properties: the straight sections of nanotubes have armchair chirality and zigzag chirality at the junctions. The diameter of the nanotubes is determined by the chirality indices (*n,m*) as $2r_0 = \frac{\sqrt{3}d_0}{\pi}\sqrt{n^2 + m^2 + nm}$ [37].

The design of the logic gate NOT includes two nanoplasmonic interferometers of the Mach-Zehnder type (figure 3(a)) [10]. The operating principle of the logic gate NOT is based on destructive interference with the simultaneous input of two pulse signals into ports *A* and *B* of the logic gate: the clock pulse and the signal pulse (the last one corresponds to the logical unit). The operating principle of the logic gate OR (figure 3(b)) is based on the unimpeded passage of pulse signals received at port *A* or port *B* of the Y-splitter [10].

In modeling process the cross-linking of two types of nanotubes "armchair" and "zigzag" has been used for the bending and splitting of the arms of logic gates [10]. This makes it possible to achieve minimal distortion of the hexagonal structure at bending angles of 150 degrees and splitting angles of 30 and 60 degrees. For coordination in places of the greatest bend, "pentagons" are used for external corners and "heptagons" and "octagons" for internal corners (figure 4).

Let us consider in more detail the operation of nanointerferometers in the NOT logic gate (figure 3(a)). We assume that the plasmon-polariton pulse at the input to port *A* or port *B* of the logic gate has the Gaussian envelope $A(t) = a_0\ exp(-t^2 T_0^{-2})\ exp(-i\omega_0 t)$, where $\omega_0$ is the carrier frequency, $T_0$ is the pulse duration, $a_0 = const$. At the distance $L$ along the $z$ axis from the input to the port, each harmonic of the signal acquires the phase delay $\beta L$ and attenuation $\alpha L$, where $\alpha$ is the attenuation coefficient, $\beta$ is the propagation constant.

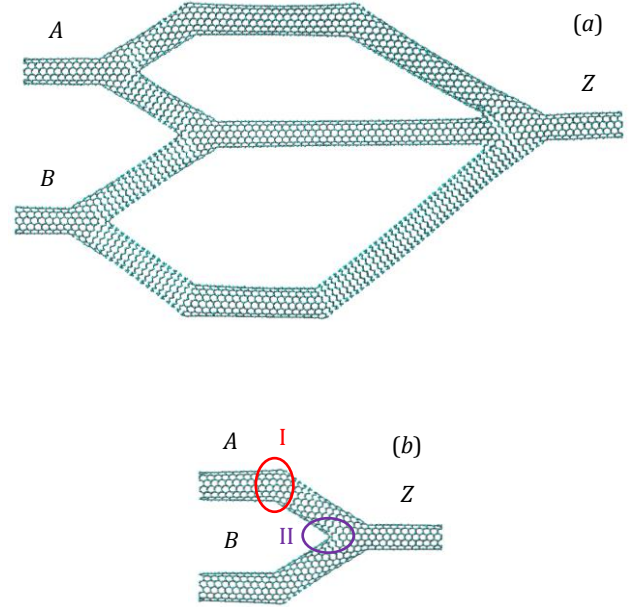

**Figure 3.** Plasmonic logic gates based on CNT: (a) NOT, (b) OR; input ports are *A* and *B*, output port is *Z*.

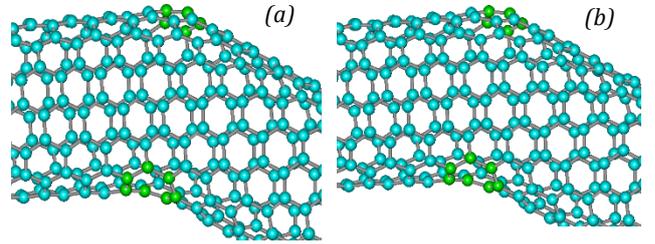

**Figure 4.** Coordination of the CNT structure in different areas: (a) 150 degrees bending (region I in figure 3); (b) 60 degrees splitting (region II in figure 3).

The pulse over the length *L* has the form [48] (Supplementary material)

$$A(t,L) = \frac{a_0}{2}\frac{\exp(-\alpha L)}{(1+\bar{L}^2)^{1/4}}\exp\left[-\frac{(t-\beta' L)^2}{T_0^2(1+\bar{L}^2)}\right]\cos\left[\omega_0 t - \beta_0 L + \frac{(t-\beta' L)^2\bar{L}}{T_0^2(1+\bar{L}^2)} - \arctan(\bar{L})\right], \qquad (23)$$

where $\bar{L} = 2\beta'' L T_0^{-2}$ is the dispersion length, $v_g = 1/\beta'$ is the group velocity, and the primes denote the frequency derivatives. As it follows from expression (23), the amplitude of the Gaussian pulse decreases by $\exp(-\alpha L)(1+\bar{L}^2)^{-1/4}$ times, and the pulse duration increases to $T = T_0\sqrt{1+\bar{L}^2}$ over the length *L*. The pulse phase acquires the time modulation $\phi_t = \frac{(t-\beta' L)^2\bar{L}}{T_0^2(1+\bar{L}^2)}$ and the shift $\phi_L = \arctan(\bar{L})$ over the length *L*.

On the input port *B* of the logic gate NOT (figure 3(a)) the clock pulse is applied at the optical carrier frequency $\omega$. The signal at the input port *B* of the nanointerferometer is divided by 50% and propagates further into its arms. With the superposition of two signals, that have passed along the arms of the nanointerferometer *B* of the same length, the constructive interference takes place, and the logical unit appears at the output port *Z*.

The length of the arms of nanointerferometer *A* is also the same. So, there is the constructive interference of signals at its output port when the signal of logical unit is applied to the input port *A*. The length of the arms of nanointerferometers *A* and *B* is different. The superposition of signals that have passed through nanointerferometers *A* and *B* leads to the interference of signals at the output port *Z*, which depends on the phase difference at the outputs of nanointerferometers $A_B = A_A \cos(\omega t - \phi_A) + A_B \cos(\omega t - \phi_B)$`

Let us represent the total signal at the output of the gate NOT in the form (Supplementary material)
$$A_{AB} = A_0 \cos(\omega t - \phi), \quad (24)$$
where $A_0 = [A_A^2 + A_B^2 + 2A_A A_B \cos(\phi_A - \phi_B)]^{\frac{1}{2}}$, $\phi = \arctan\left[\frac{A_A \sin(\phi_A) + A_B \sin(\phi_B)}{A_A \cos(\phi_A) + A_B \cos(\phi_B)}\right]$. The destructive interference of the signal pulse and the clock pulse at the output of the logic gate NOT is observed when the phase difference between the signals passed through the nanointerferometers *A* and *B* has the form
$$\phi_A - \phi_B = \beta_0(L_A - L_B) + \arctan(\overline{L}_A) - \arctan(\overline{L}_B) = (2j+1)\pi, \quad (25)$$
where $j = 0,1,2,...$. In this case, there is no signal at the port *Z* and the logical zero appears at the output port of the gate when the difference in the lengths of the arms of nanointerferometers *A* and *B* is equal to $L_A - L_B \approx (2j+1)\lambda/2$. For the logic gate NOT, the ratio of the amplitudes of the signal pulse and the clock pulse, as well as the phase shift between them, must ensure the condition of visibility of their interference $(I_{max} - I_{min})/(I_{max} + I_{min}) \geq \frac{1}{3}$, where $I = A_0^2$ is the intensity of signal.

We have estimated the loss coefficient of SPP pulse in CNT as $\alpha = 1/l_p$, where $l_p \sim v_g \tau_p$ is the free path of SPP, $\tau_p = 3 \cdot 10^{-12} s$ is the SPP lifetime [22]. Assuming $v_g \cong v = \omega/\beta = 2.52 \cdot 10^{10}$ cm/s, we have obtained the value of the loss coefficients $\alpha\_TM = 154$ cm$^{-1}$, $\alpha\_TE = 124$ cm$^{-1}$, and the insertion loss $IL_{TM} = 0.10$ dB and $IL_{TE} = 0.083$ dB in the logic gate NOT.

A review of the all optical logic gates including plasmonic logic gates is presented in the article [49]. The CNT logic gates NOT and OR are seamlessly coupled with the CNT nanolaser proposed in our work. The signal from the nanolaser or another logic gate enters to the logic gate NOT or OR and is processed by it. The CNT nanolaser can also be used to amplify the signal coming from another logic gate. Thus, the plasmonic circuit on carbon nanotubes operates at an optical frequency without transforming signals from optical to electrical and vice a verse. The operating frequency of such plasmonic circuitry is tens of terahertz. The CNT logic gates NOT and OR and the CNT nanolaser provide the complete element base for processor circuitry.

## 14. Conclusion

Optical pumping of the nanoresonator on CNT with zigzag chirality with semiconductor properties increases the concentration of electrons in the conduction band. This process creates conditions for the nanolaser to generate the flow of plasmon-polaritons when the gain value exceeds the generation threshold for the nanoresonator with the Bragg gratings in the form of periodic inhomogeneities at the ends of the CNT. Thus, when the gain value exceeds the threshold, the nanolaser on zigzag CNT generates the flow of coherent plasmon-polaritons, which can be output into a CNT plasmonic waveguide. The length of the nanoresonator determines the wavelength of coherent plasmon-polaritons. In particular, it is possible to generate the plasmon-polaritons in the nanolaser on CNT at the telecommunication frequency corresponding to the wavelength of 1.55 $\mu m$ in air.

The advantage of the proposed CNT nanolaser is the direction of generated plasmonic pulse into the attached CNT plasmonic waveguide, i.e. no focusing system for the generated radiation by the CNT nanolaser is needed.

Based on the theoretical analysis of signal propagation in CNTs, we propose the models of plasmonic logic gates NOT and OR representing a complete functional basis for binary logic. The plasmonic logic gate NOT consists of two nanointerferometers of the Mach-Zehnder type based on CNTs, and the plasmonic logic gate OR is implemented on the basis of Y-splitter on CNTs.

## Data availability statement

All data that support the findings of this study are included within the article and supplementary document.

## Acknowledgments

The authors are grateful for support to the Russian Science Foundation (RSF) (19-72-20154).

## Conflict of interest

The authors declare no conflict of interests.

## ORCID iDs

I V Dzedolik https://orcid.org/0000-0003-2761-4611
S V Tomilin https://orcid.org/0000-0002-0668-0647

# NANOLASER AND LOGIC GATES ON CARBON NANOTUBES: SUPPLEMENTARY MATERIAL

## 1. Dispersion equation

The expressions for the mode components in a carbon nanotube (CNT) have the form of Macdonald functions [1]

$$E_r = \frac{r_0^2}{w^2}\left[A\frac{-i\beta w}{r_0 K_l(w)}K'_l\left(\frac{wr}{r_0}\right) + B\frac{k_0}{K_l(w)}\frac{l}{r}K_l\left(\frac{wr}{r_0}\right)\right],$$
$$E_\varphi = \frac{r_0^2}{w^2}\left[A\frac{\beta}{K_l(w)}\frac{l}{r}K_l\left(\frac{wr}{r_0}\right) + B\frac{ik_0 w}{r_0 K_l(w)}K'_l\left(\frac{wr}{r_0}\right)\right],$$
$$H_r = \frac{r_0^2}{w^2}\left[B\frac{-i\beta w}{r_0 K_l(w)}K'_l\left(\frac{wr}{r_0}\right) - A\frac{k_0\varepsilon}{K_l(w)}\frac{l}{r}K_l\left(\frac{wr}{r_0}\right)\right], \qquad (S.1)$$
$$H_\varphi = \frac{r_0^2}{w^2}\left[B\frac{\beta}{K_l(w)}\frac{l}{r}K_l\left(\frac{wr}{r_0}\right) - A\frac{ik_0\varepsilon w}{r_0 K_l(w)}K'_l\left(\frac{wr}{r_0}\right)\right],$$
$$E_z = A\frac{K_l(wr/r_0)}{K_l(w)}, \quad H_z = B\frac{K_l(wr/r_0)}{K_l(w)},$$

where the prime denotes the derivative of the function with respect to its argument, $w^2 = r_0^2(\beta^2 - k_0^2\varepsilon)$, $k_0 = \omega/c$. The dependence of the modes on time, azimuthal and longitudinal coordinates has the form $E_j, H_j \sim \exp(-i\omega t + il\varphi + i\beta z)$ $\quad l = 0, \pm 1, \pm 2, \ldots$, $\varepsilon$ is the permittivity of the medium surrounding the CNT.

Substituting into the Leontovich boundary conditions [2] $E_z = \zeta H_\varphi$ and $H_z = E_\varphi/\zeta$ on the nanotube surface at $r = r_0$ the expressions for the mode components (S.1) $AK_l(w) = \zeta\frac{r_0}{w^2}[Bl\beta K_l(w) - Aik_0\varepsilon w K'_l(w)]$, $BK_l(w) = \frac{r_0}{\zeta w^2}[Al\beta K_l(w) + Bik_0 w K'_l(w)]$, we obtain the system of homogeneous linear equations for the amplitudes $A$ and $B$

$$\left[K_l(w) + i\zeta\frac{\varepsilon k_0 r_0}{w}K'_l(w)\right]A - \left[\zeta\frac{l\beta r_0}{w^2}K_l(w)\right]B = 0,$$
$$\left[\frac{l\beta r_0}{\zeta w^2}K_l(w)\right]A + \left[i\frac{k_0 r_0}{\zeta w}K'_l(w) - K_l(w)\right]B = 0, \qquad (S.2)$$

where the impedance for the CNT with $\mu_c = 1$ is equal to $\zeta = 1/\sqrt{\varepsilon_c}$. We find the determinant of the system of equations (S.2)

$$D = -K_l^2(w) - i\zeta\frac{\varepsilon k_0 r_0}{w}K_l(w)K'_l(w) + i\frac{k_0 r_0}{\zeta w}K_l(w)K'_l(w) - \varepsilon\frac{k_0^2 r_0^2}{w^2}K'^2_l(w) + l^2\frac{\beta^2 r_0^2}{w^4}K_l^2(w). \qquad (S.3)$$

Equating the determinant (S.3) to zero, we obtain the equation

$$K_l^2(w) + \varepsilon\frac{k_0^2 r_0^2}{w^2}K'^2_l(w) + i\left(\zeta\varepsilon - \frac{1}{\zeta}\right)\frac{k_0 r_0}{w}K_l(w)K'_l(w) = l^2\frac{\beta^2 r_0^2}{w^4}K_l^2(w). \qquad (S.4)$$

The coefficient in front of the third term of Eq. (S.4) is equal to $\zeta\varepsilon - \frac{1}{\zeta} = \frac{\varepsilon - \varepsilon_c}{\sqrt{\varepsilon_c}}$. Then from Eq. (S.4) we obtain the dispersion equation for the propagation constants $\beta_l(\omega)$ of the plasmon-polariton modes of the CNT in the form

$$K_l^2(w) + \varepsilon\frac{k_0^2 r_0^2}{w^2}K'^2_l(w) + i\frac{\varepsilon - \varepsilon_c}{\sqrt{\varepsilon_c}}\frac{k_0 r_0}{w}K_l(w)K'_l(w) = l^2\frac{\beta^2 r_0^2}{w^4}K_l^2(w). \qquad (S.5)$$

We obtain the dispersion equation from equation (S.5) for modes with zero azimuthal index $l = 0$ taking into account the property of the Macdonald function $K'_0(w) = -K_1(w)$,

$$K_0^2(w) + \varepsilon\frac{k_0^2 r_0^2}{w^2}K_1^2(w) - i\frac{\varepsilon - \varepsilon_c}{\sqrt{\varepsilon_c}}\frac{k_0 r_0}{w}K_0(w)K_1(w) = 0. \qquad (S.6)$$

Presenting equation (S.6) in the form $\frac{K_1^2(w)}{K_0^2(w)} - i\frac{\varepsilon - \varepsilon_c}{\varepsilon\sqrt{\varepsilon_c}}\frac{w}{k_0 r_0}\frac{K_1(w)}{K_0(w)} + \frac{w^2}{\varepsilon k_0^2 r_0^2} = 0$ and denoting $K_1(w)K_0^{-1}(w) = \gamma$, we look for the roots of the quadratic equation $\gamma^2 - ia_1\gamma + a_2 = 0$, where

$a_1 = \frac{\varepsilon - \varepsilon_c}{\varepsilon\sqrt{\varepsilon_c}} \frac{w}{k_0 r_0}$, $a_2 = \frac{w^2}{\varepsilon k_0^2 r_0^2}$, we get $\gamma_{1,2} = i\left(\frac{a_1}{2} \pm \sqrt{\frac{a_1^2}{4} + a_2}\right)$. From here we find $\gamma_{1,2} = i\frac{w}{k_0 r_0}\left(\frac{\varepsilon - \varepsilon_c}{2\varepsilon\sqrt{\varepsilon_c}} \pm \sqrt{\frac{(\varepsilon - \varepsilon_c)^2}{4\varepsilon^2 \varepsilon_c} + \frac{1}{\varepsilon}}\right)$, i.e. the roots of the equation are $\gamma_1 = i\frac{w}{\sqrt{\varepsilon_c} k_0 r_0}$ and $\gamma_2 = -i\frac{w\sqrt{\varepsilon_c}}{\varepsilon k_0 r_0}$.

## 2. Reflection coefficients of nanoresonator

The continuity conditions [3] for the tangential components of TM mode $H_\varphi = \sqrt{\varepsilon_c}A$, $E_z = A$ that incident on the inhomogeneity in the nanoresonator (i - incident, r - reflected, t - transmitted) have the form
$$E_{ri} + E_{rr} = E_{rt}, \quad H_{\varphi i} - H_{\varphi r} = H_{\varphi t}. \tag{S.7}$$
The continuity conditions for TE mode have the form
$$H_{ri} + H_{rr} = H_{rt}, \quad E_{\varphi i} - E_{\varphi r} = E_{\varphi t}. \tag{S.8}$$
We find the connection between the components of the TM mode $H_\varphi = \frac{\varepsilon k_0}{\beta_0} E_r$ and the TE mode $H_r = -\frac{\beta_0}{k_0} E_\varphi$. The continuity conditions for the forward (+) and backward (-) modes inside the inhomogeneity are $E = E^+ + E^-$ and $H = H^+ + H^-$. Then we obtain the relations for the TM mode $E_h = E_{ri} + E_{rr}$, $H_h = \frac{\varepsilon k_0}{\beta_{0h}}(E_{ri} - E_{rr})$, and for the TE mode $H_h = \left(-\frac{\beta_0}{k_0}\right)(E_{\varphi i} + E_{\varphi r})$, $E_h = E_{\varphi i} - E_{\varphi r}$.

These expressions have the form:
for TM mode
$$E_h = Ae^{i\beta_{0h}y} + Be^{-i\beta_{0h}y}, \quad H_h = \frac{\varepsilon k_0}{\beta_{0h}}\left(Ae^{i\beta_{0h}y} - Be^{-i\beta_{0h}y}\right), \tag{S.9}$$
for TE mode
$$E_h = Ae^{i\beta_{0h}y} - Be^{-i\beta_{0h}y}, \quad H_h = \left(-\frac{\beta_0}{k_0}\right)\left(Ae^{i\beta_{0h}y} + Be^{-i\beta_{0h}y}\right). \tag{S.10}$$

Taking into account the continuity of the tangential components of the modes on the left $y = 0$ and right $y = d$ boundaries of one inhomogeneity in the CNT, from expressions (S.9) and (S.10) we obtain systems of equations. Excluding the amplitudes $A$ and $B$ from the obtained equations, we find unimodular matrices for TM mode
$$\begin{pmatrix} E_0 \\ H_0 \end{pmatrix} = \begin{pmatrix} \cos(\beta_{0h}d) & -i\frac{\beta_{0h}}{\varepsilon k_0}\sin(\beta_{0h}d) \\ -i\frac{\varepsilon k_0}{\beta_{0h}}\sin(\beta_{0h}d) & \cos(\beta_{0h}d) \end{pmatrix} \begin{pmatrix} E_d \\ H_d \end{pmatrix} = M_{TM1} \begin{pmatrix} E_d \\ H_d \end{pmatrix}, \tag{S.11}$$
and for TE mode
$$\begin{pmatrix} E_0 \\ H_0 \end{pmatrix} = \begin{pmatrix} \cos(\beta_{0h}d) & i\frac{k_0}{\beta_0}\sin(\beta_{0h}d) \\ i\frac{\beta_0}{k_0}\sin(\beta_{0h}d) & \cos(\beta_{0h}d) \end{pmatrix} \begin{pmatrix} E_d \\ H_d \end{pmatrix} = M_{TE1} \begin{pmatrix} E_d \\ H_d \end{pmatrix}. \tag{S.12}$$

For $N$ inhomogeneities with width $d_j$ along the coordinate $y = d_1 + d_2 + \cdots + d_N$ we obtain in the last plane $N+1$ the expression
$$\begin{pmatrix} E_0 \\ H_0 \end{pmatrix} = M_1 M_2 \ldots M_N \begin{pmatrix} E_{N+1} \\ H_{N+1} \end{pmatrix}. \tag{S.13}$$
The product of $N$ unimodular matrices characterizes the Bragg grating
$$M = \begin{pmatrix} m_{11} & m_{12} \\ m_{21} & m_{22} \end{pmatrix} = \prod_{j=1}^{N} M_j. \tag{S.14}$$
We find the elements $m_{ij}$ of the matrix $M$ using the expressions for the unimodular matrix for the TM mode (S.9) and the matrix for the TE mode (S.10), and obtain the equations for the TM mode
$$E_{ri} + E_{rr} = \left(m_{11} + m_{12}\frac{\varepsilon k_0}{\beta_0}\right)E_{rt}, \quad E_{ri} - E_{rr} = \left(\frac{\beta_0}{\varepsilon k_0}m_{21} + m_{22}\right)E_{rt}, \tag{S.15}$$
$$E_{\varphi i} - E_{\varphi r} = \left(m_{11} - \frac{\beta_0}{k_0}m_{12}\right)E_{\varphi t}, \quad E_{\varphi i} + E_{\varphi r} = \left(m_{22} - \frac{k_0}{\beta_0}m_{21}\right)E_{\varphi t}. \tag{S.16}$$

From equations (S.15) and (S.16) we find the amplitude coefficients of transmission $\tau$ and reflection $\rho$ for TM mode

$$\tau = \frac{E_{rt}}{E_{ri}} = \frac{2}{m_{11}+m_{12}\frac{\varepsilon k_0}{\beta_0}+\frac{\beta_0}{\varepsilon k_0}m_{21}+m_{22}}, \quad \rho = \frac{E_{rr}}{E_{ri}} = \frac{m_{11}+m_{12}\frac{\varepsilon k_0}{\beta_0}-\frac{\beta_0}{\varepsilon k_0}m_{21}-m_{22}}{m_{11}+m_{12}\frac{\varepsilon k_0}{\beta_0}+\frac{\beta_0}{\varepsilon k_0}m_{21}+m_{22}}, \tag{S.17}$$

and for TE mode

$$\tau = \frac{E_{\varphi t}}{E_{\varphi i}} = \frac{2}{m_{11}-\frac{\beta_0}{k_0}m_{12}-\frac{k_0}{\beta_0}m_{21}+m_{22}}, \quad \rho = \frac{E_{\varphi r}}{E_{\varphi i}} = \frac{m_{22}-\frac{k_0}{\beta_0}m_{21}-m_{11}+\frac{\beta_0}{k_0}m_{12}}{m_{11}-\frac{\beta_0}{k_0}m_{12}-\frac{k_0}{\beta_0}m_{21}+m_{22}}. \tag{S.18}$$

## 3. Permittivity of carbon nanotube

Taking into account the relaxation frequencies (spontaneous emission) for electron transitions, the average dipole moment $\langle \mathbf{d}_1 \rangle$ of an electron [4] is equal to

$$\langle \mathbf{d}_1 \rangle = \frac{e^2 \mathbf{E}_a}{2\hbar}\sum_n |r_{n0}|^2 \left( \frac{e^{i\omega t}}{\omega_{n0}-i\Gamma_n/2+\omega} + \frac{e^{-i\omega t}}{\omega_{n0}-i\Gamma_n/2-\omega} + \frac{e^{i\omega t}}{\omega_{n0}-i\Gamma_n/2-\omega} + \frac{e^{-i\omega t}}{\omega_{n0}-i\Gamma_n/2+\omega} \right), \tag{S.19}$$

and we rewrite it in the form

$$\langle \mathbf{d}_1 \rangle = \frac{e^2 \mathbf{E}_a \cos(\omega t)}{\hbar}\sum_n |r_{n0}|^2 \left( \frac{1}{\omega_{n0}-i\Gamma_n/2+\omega} + \frac{1}{\omega_{n0}-i\Gamma_n/2-\omega} \right). \tag{S.20}$$

Transforming expression (S.20), we obtain the electron dipole moment taking into account resonant absorption and spontaneous emission in CNT

$$\langle \mathbf{d}_1 \rangle = \frac{2e^2 \mathbf{E}_a \cos(\omega t)}{\hbar}\sum_n |r_{n0}|^2 \frac{\omega_{n0}-i\Gamma_n/2}{(\omega_{n0}-i\Gamma_n/2)^2-\omega^2}. \tag{S.21}$$

We find the permittivity of CNT using the expression for the induction vector $\mathbf{D} = \mathbf{E} + 4\pi \mathbf{P} = \varepsilon \mathbf{E}$, substituting the average electron dipole moment (S.2) into the polarization vector of the medium $\mathbf{P} = \langle \mathbf{d}_1 \rangle N_e$, where $N_e$ is the number of electrons per unit volume,

$$\varepsilon_c(\omega) = 1 + \frac{2m\omega_e^2}{\hbar}\sum_n |r_{n0}|^2 \frac{\omega_{n0}-i\Gamma_n/2}{(\omega_{n0}-i\Gamma_n/2)^2-\omega^2}. \tag{S.22}$$

Here $\omega_e^2 = 4\pi e^2 N_e/m_{eff}$ is the square of electron plasma frequency, $m_{eff}$ is the effective mass of electron.

We assume that plasmon-polariton modes are excited at the frequency of the direct electron transition $\omega = \omega_{nn'}$ in CNT. Then the electron dipole moment (S.20) is equal to

$$\langle \mathbf{d}_1 \rangle = \frac{2\mathbf{E}_a \cos(\omega t) e^2 |r_{nn'}|^2}{\hbar}\left( \frac{1}{4\omega - i\Gamma} + \frac{i}{\Gamma} \right). \tag{S.23}$$

Substituting the electron dipole moment (S.23) into the polarization vector of the medium, we obtain the expression for the permittivity of CNT upon emission (absorption) of the photon at the frequency of electron transition $\omega$,

$$\varepsilon_c(\omega) = 1 + \frac{8\pi e^2 |r_{nn'}|^2 N_e}{\hbar}\left[ \frac{4\omega}{16\omega^2+\Gamma^2} + \frac{i}{\Gamma}\left(1 + \frac{\Gamma^2}{16\omega^2+\Gamma^2}\right) \right]. \tag{S.24}$$

Let us find the relaxation frequency $\Gamma_n$ as the reciprocal of the electron lifetime $\tau_n = 1/A_{n0}$ in the conduction band before the direct transition to the valence band. We determine the energy density emitted by the electron during a dipole transition per unit time by integrating the Poynting vector with the modulus $S = \frac{c}{4\pi}\frac{(d^r)^2}{r^2}\frac{\omega^4}{c^4}\sin^2\theta$ at $\omega r/c \ll 1$ over the surface with radius $r = a$ surrounding the dipole $d^r = -er_{n0}\exp(-i\omega t) + c.c.$ [6]. Taking into account the values of the time-averaged square of the dipole moment $\overline{(d^r)^2} = \overline{(2d_0^r \cos\omega t)^2} = 2(d_0^r)^2$ and the integral over the surface $r^2\int_0^{2\pi}d\varphi\int_0^\pi S\sin\theta d\theta = \frac{4\pi r^2}{3}\frac{c}{4\pi}\frac{(d^r)^2}{r^2}\frac{\omega^4}{c^4}$, we find the energy density of the dipole radiation per unit time $E_1 = \frac{2e^2\omega_{n0}^4}{3c^3}(r_{n0})^2$.

Let us estimate the coefficient before the square brackets in expression (S.24) for the direct transition of an electron during recombination from the conduction band to the valence band.

The energy density of dipole radiation per unit time is $E_1$. The Einstein coefficient [5,6] for spontaneous oscillator emission $A_{n0}$ is equal to the oscillator radiation energy per unit time divided by the photon energy $\hbar\omega_{n0}$, i.e. the relaxation frequency is $\Gamma_n = A_{n0} = \frac{2e^2\omega_{n0}^3}{3\hbar c^3}|r_{n0}|^2$. Assuming that the relation $e^2/r_{n0} = \hbar\omega_{n0}$ is satisfied for the electron during the direct transition from the conduction band to the valence band, we find the relaxation frequency $\Gamma_n = \frac{2e^6\omega_{n0}}{3\hbar^3 c^3}$. Then the coefficient in expression (S.24) takes the form $\frac{8\pi e^2 |r_{nn'}|^2 N_e}{\hbar} = \frac{8\pi e^6 N_e}{\hbar^3 \omega^2}$. Substituting the expression for $\Gamma$ into the coefficient $\frac{8\pi e^6 N_e}{\hbar^3 \omega^2} = \frac{12\pi c^3 N_e}{\omega^3}\Gamma$, we obtain the expression for the permittivity of the CNT at the transition frequency $\omega$ in the form

$$\varepsilon_c(\omega) = 1 + 12\pi c^3 N_e \left[\frac{4\Gamma}{\omega^2(16\omega^2+\Gamma^2)} + \frac{i}{\omega^3}\left(1 + \frac{\Gamma^2}{16\omega^2+\Gamma^2}\right)\right]. \tag{S.25}$$

## 4. Number of interband electron transitions

Number of electronic transitions [5] $N = \int_0^\infty dk\, w_{cv}\rho(k)$ is equal to $N = \frac{2V}{\hbar\pi}\int_0^\infty dk\, k^2 |\hat{H}'_{vc}|^2 \delta\left(\frac{\hbar^2 k^2}{2m_{vc}} + E_g - \hbar\omega\right)$ for a semiconductor with volume $V$ at probability $w_{vc} = \frac{2\pi}{\hbar}|\hat{H}'_{vc}|^2 \delta\left(\frac{\hbar^2 k^2}{2m_{vc}} + E_g - \hbar\omega\right)$ and density of electronic states $\rho(k) = 2 \cdot \frac{4\pi k^2}{8\pi^3}V = \frac{k^2 V}{\pi^2}$. By introducing the variable $X = \frac{\hbar^2 k^2}{2m_{vc}} + E_g - \hbar\omega$, we find $k = \sqrt{\frac{2m_{vc}}{\hbar^2}(X - E_g + \hbar\omega)}$ and obtain the number of transitions $N = \frac{2V}{\hbar\pi}\int_0^\infty dX\, \frac{m_{vc}}{\hbar^2}\sqrt{\frac{2m_{vc}}{\hbar^2}(X - E_g + \hbar\omega)}|\hat{H}'_{vc}|^2 \delta(X)$. At $X=0$, i.e. at $\frac{\hbar^2 k^2}{2m_{vc}} + E_g = \hbar\omega$, the number of interband transitions in CNT is

$$N = \frac{V(2m_{vc})^{3/2}}{\hbar^4 \pi}|\hat{H}'_{vc}|^2 \sqrt{\hbar\omega - E_g}. \tag{S.26}$$

## 5. Matrix elements of the perturbation operator

The Poynting vector of the TM mode ($E_r, H_\varphi, E_z$) is equal to $\mathbf{S} = \frac{c}{4\pi}\left[\mathbf{1}_r(-E_z H_\varphi) + \mathbf{1}_z(E_r H_\varphi)\right]$. The time-averaged Poynting vector for real values of the mode components is equal to

$$\bar{\mathbf{S}} = \frac{c}{8\pi}\left[-\mathbf{1}_r(E_z H_\varphi^* + H_\varphi E_z^*) + \mathbf{1}_z(E_r H_\varphi^* + H_\varphi E_r^*)\right]. \tag{S.27}$$

On the surface of the CNT at $r = r_0$ we find the modulus of the Poynting vector for the TM mode in the form

$$\bar{S}_{TM} = \frac{c}{8\pi}\left[(E_z H_\varphi^* + H_\varphi E_z^*)^2 + (E_r H_\varphi^* + H_\varphi E_r^*)^2\right]^{1/2} = \frac{c}{4\pi}S_{TM}A^2, \tag{S.28}$$

where $S_{TM} = \sqrt{\beta_0^2/k_0^2 + \varepsilon}$.

The Poynting vector of the TE mode ($H_r, E_\varphi, H_z$) is equal to $\mathbf{S} = \frac{c}{4\pi}\left[\mathbf{1}_r(E_\varphi H_z) + \mathbf{1}_z(-E_\varphi H_r)\right]$, and, averaging over time, we obtain $\bar{\mathbf{S}} = \frac{c}{8\pi}\left[\mathbf{1}_r(E_\varphi H_z^* + H_z E_\varphi^*) - \mathbf{1}_z(E_\varphi H_r^* + H_r E_\varphi^*)\right]$, i.e.

$$\bar{S}_{TE} = \frac{c}{8\pi}\left[(E_\varphi H_z^* + H_z E_\varphi^*)^2 + (E_\varphi H_r^* + H_r E_\varphi^*)^2\right]^{1/2} = \frac{c}{4\pi}S_{TE}B^2, \tag{S.29}$$

where $S_{TE} = \varepsilon^{-1}\sqrt{\beta_0^2/k_0^2 + \varepsilon}$.

Matrix element of the perturbation Hamiltonian on the CNT surface at $r = r_0$ for TM mode is

$$\hat{H}'_{vc} = \frac{eA}{2}\int dz\, \psi_v^*(y)z\psi_c(y) = \frac{eE_z}{2}y_{vc}, \tag{S.30}$$

for TE mode is

$$\hat{H}'_{vc} = \frac{er_0 B}{2}\int d\varphi\, \psi_v^*(\varphi)\varphi\psi_c(\varphi) = \frac{er_0 E_\varphi}{2}\varphi_{vc}. \tag{S.31}$$

squares of matrix element for
TM mode is

$$|\hat{H}'_{vc}|^2 = \frac{e^2 E_z^2}{4}|y_{vc}|^2 = \frac{e^2 A^2}{4}|y_{vc}|^2, \tag{S.32}$$

for TE mode is

$$|\hat{H}'_{vc}|^2 = \frac{e^2 r_0^2 E_\varphi^2}{4}|\varphi_{vc}|^2 = \frac{e^2 r_0^2 B^2}{4\varepsilon}|\varphi_{vc}|^2. \tag{S.33}$$

## 6. Gaussian pulse

At the entrance to the port, the plasmonic pulse is represented in the form

$$A(t) = a(t)\exp(-i\omega_0 t), \tag{S.34}$$

where $a(t)$ is the envelope of the wave packet, $\omega_0$ is the carrier frequency. We find the spectral form of the wave packet using the integral Fourier transform

$$A(\omega) = \int_{-\infty}^{\infty} a(t)\exp(-i\omega_0 t)\exp(i\omega t)dt, \tag{S.35}$$

assuming that the envelope $a(t)$ differs from zero only in a limited time interval of the order of the pulse duration $\sim T_0$. The spectral form of a Gaussian pulse $A(t) = a_0 \exp(-t^2 T_0^{-2})\exp(-i\omega_0 t)$ has the form

$$A(\omega) = a_0 \int_{-\infty}^{\infty} \exp(i(\omega - \omega_0)t - t^2 T_0^{-2})\, dt. \tag{S.36}$$

The integral (S.36) in frequency space is limited, the pulse that has propagated over the distance $L$ in a linear medium $A(\omega, L) = A(\omega)\exp[-(\alpha - i\beta)L]$ has the form

$$A(t, L) = \frac{1}{2\pi}\int_{-\infty}^{\infty} A(\omega)\exp[-(\alpha - i\beta)L]\, d\omega. \tag{S.37}$$

Let us assume that the pulse envelope $a(t) = a_0 \exp(-t^2 T_0^{-2})$ changes slowly compared to the oscillation rate of the carrier harmonic $\omega_0$ [7]. Then the dependence of the propagation constant $\beta$ on the frequency is given by the dispersion law for the given medium $\beta = \beta(\omega)$ for the relatively narrow spectral interval $\Delta\omega = \omega - \omega_0$ of the signal, the propagation constant $\beta$ can be expanded in the Taylor series in the vicinity of the carrier frequency $\beta = \beta_0 + \frac{d\beta}{d\omega}|_{\omega_0}\Delta\omega + \frac{1}{2}\frac{d^2\beta}{d\omega^2}|_{\omega_0}\Delta\omega^2 + \cdots$. Substituting this series into integral (S.37) up to the second derivative of the propagation constant, we obtain the pulse shape over the length $L$

$$A(t, L) = \frac{\exp[-(\alpha - i\beta_0)L - i\omega_0 t]}{2\pi}\int_{-\infty}^{\infty} A(\omega)\exp\left\{-i\left[(t - \beta' L)\Delta\omega + \frac{1}{2}\beta''\Delta\omega^2 L\right]\right\}d\Delta\omega, \tag{S.38}$$

where $\beta' = (d\beta/d\omega)_{\omega_0}$ and $\beta'' = (d^2\beta/d\omega^2)_{\omega_0}$. It follows from expression (S.38) that the pulse decreases in amplitude by $\exp(-\alpha L)$ times, and its phase delay is equal to $L/v_g$, where $v_g = 1/\beta'$ is the group velocity. The term in the phase proportional to $\beta''$ characterizes the distortion of the signal as a result of the dispersion of the group velocity $\beta'' = -v_g^{-2} dv_g/d\omega$.

Let us single out in expression (S.36) $A(\omega) = a_0 \int_{-\infty}^{\infty} \exp(i(\omega - \omega_0)t - t^2 T_0^{-2})\, dt$ the perfect square in the exponent $a_1 t^2 + a_2 t + a_3 = (b_1 t + b_2)^2 = b_1^2 t^2 + 2 b_1 b_2 t + b_2^2$, where $a_1 = T_0^{-2}$, $a_2 = -i(\omega - \omega_0)$, and $b_1 = \sqrt{a_1} = 1/T_0$, $b_2 = a_2/2b_1 = -i(\omega - \omega_0)T_0/2$, $a_3 = b_2^2 = -(\omega - \omega_0)^2 T_0^2/4$. Then we obtain the spectral shape of the pulse, i.e. the form of Gaussian plasmon pulse in the energy space $\omega$

$$A(\omega) = a_0 \exp(a_3)\frac{1}{b_1}\int_{-\infty}^{\infty} \exp(-x^2)dx = a_0 T_0 \sqrt{\pi}\exp\left(-\frac{(\omega - \omega_0)^2 T_0^2}{4}\right). \tag{S.39}$$

We obtain the pulse shape at length $L$ by substituting expression (S.39) into expression (S.38)

$$A(t, L) = \frac{a_0 T_0 \sqrt{\pi}}{2\pi} e^{-\alpha L}\exp(-i\omega_0 t + i\beta_0 L)\int_{-\infty}^{\infty} \exp\left[-\frac{\Delta\omega^2 T_0^2}{4} - i(t - \beta' L)\Delta\omega + i\frac{\beta'' L}{2}\Delta\omega^2\right]d\Delta\omega. \tag{S.40}$$

The perfect square in the exponent is $a_1 = T_0^2/4 - i\beta''L/2$, $a_2 = i(t - \beta'L)$, $b_1 = \sqrt{a_1} = \sqrt{T_0^2/4 - i\beta''L/2}$, $b_2 = \frac{a_2}{2b_1} = \frac{i(t-\beta'L)}{2\sqrt{T_0^2/4 - i\beta''L/2}}$, $a_3 = b_2^2 = -(t - \beta'L)^2/(T_0^2 - i2\beta''L)$. By integrating, we find the shape of the plasmon pulse in two-dimensional space

$$A(t,L) = \frac{a_0}{2}\frac{\exp(-\alpha L)}{\sqrt{1-i\bar{L}}} \exp\left[-\frac{(t-\beta'L)^2}{T_0^2(1-i\bar{L})}\right] \exp(-i\omega_0 t + i\beta_0 L), \tag{S.41}$$

where $\bar{L} = 2\beta''L/T_0^2$ is the dispersion length. We select the real amplitude and phase in (S.41) according to the formula $a' + ia'' = \sqrt{(a')^2 + (a'')^2} \exp[i\arctan(a''/a')]$, and obtain the expression for the pulse

$$A(t,L) = \frac{a_0}{2}\frac{\exp(-\alpha L)}{(1+\bar{L}^2)^{\frac{1}{4}}} \exp\left[-\frac{(t-\beta'L)^2}{T_0^2(1+\bar{L}^2)}\right] \exp\left\{i\left[-\omega_0 t + \beta_0 L - \frac{(t-\beta'L)^2\bar{L}}{T_0^2(1+\bar{L}^2)} + \arctan(\bar{L})\right]\right\}. \tag{S.42}$$

## 7. Carrier wave interference in the logic gate NOT

At superposition of two signals that propagated through the branches 1 and 2 of the interferometer

$$A_{12} = A_1 + A_2 = A_1\cos(\omega t - \phi_1) + A_2\cos(\omega t - \phi_2), \tag{S.43}$$

we represent the total signal in the form

$$A_{12} = A_0\cos(\omega t - \phi). \tag{S.44}$$

Then, equating (S.43) and (S.44) we obtain
$A_0\cos(\omega t)\cos(\phi) + A_0\sin(\omega t)\sin(\phi) = A_1\cos(\omega t)\cos(\phi_1) + A_1\sin(\omega t)\sin(\phi_1) + A_2\cos(\omega t)\cos(\phi_2) + A_2\sin(\omega t)\sin(\phi_2)$, where do we find it $A_0\cos(\phi) = A_1\cos(\phi_1) + A_2\cos(\phi_2)$ and $A_0\sin(\phi) = A_1\sin(\phi_1) + A_2\sin(\phi_2)$. We find the amplitude and phase in expression (S.44) from the obtained equalities

$$A_0 = [A_1^2 + A_2^2 + 2A_1A_2\cos(\phi_1 - \phi_2)]^{1/2}, \tag{S.45}$$

$$\varphi = \arctan\left[\frac{A_1\sin(\phi_1) + A_2\sin(\phi_2)}{A_1\cos(\phi_1) + A_2\cos(\phi_2)}\right]. \tag{S.46}$$